\documentclass[applemac,letter]{aa}
\usepackage{inputenc}
\usepackage{natbib}
\usepackage{txfonts}
\usepackage{graphicx}
\usepackage{color}
\bibpunct{(}{)}{;}{a}{}{,}

\begin{document}

\title{Understanding the $\rm C_3H_2$ cyclic-to-linear ratio in L1544}
\author{O. Sipil\"a,
		S. Spezzano,
	     \and{P. Caselli}
}
\institute{Max-Planck-Institute for Extraterrestrial Physics (MPE), Giessenbachstr. 1, 85748 Garching, Germany \\
e-mail: \texttt{osipila@mpe.mpg.de}
}

\date{Received / Accepted}

\abstract
{}
{We aim to understand the high cyclic-to-linear $\rm C_3H_2$ ratio ($32 \pm 4$) observed toward L1544 by \citet{Spezzano16}.}
{We combine a gas-grain chemical model with a physical model for L1544 to simulate the column densities of cyclic and linear $\rm C_3H_2$ observed toward L1544. The most important reactions for the formation and destruction of both forms of $\rm C_3H_2$ are identified, and their relative rate coefficients are varied to find the best match to the observations.}
{We find that the ratio of the rate coefficients of $\rm C_3H_3^+ + e^- \longrightarrow C_3H_2 + H$ for cyclic and linear $\rm C_3H_2$ must be $\sim 20$ in order to reproduce the observations, depending on the branching ratios assumed for the $\rm C_3H_3^+ + e^- \longrightarrow C_3H + H_2$ reaction. In current astrochemical networks it is assumed that cyclic and linear $\rm C_3H_2$ are formed in a 1:1 ratio in the aforementioned reactions. Laboratory studies and/or theoretical calculations are needed to confirm the results of our chemical modeling based on observational constraints.}
{}


\maketitle

\section{Introduction}

Isomers and isotopologs are excellent tools to probe physical and chemical properties, as well as evolutionary states, in the ISM. Cyclopropenylidene, the cyclic form of C$_3$H$_2$, is one of the most abundant and widespread molecules in our Galaxy \citep{mat85}. Propadienylidene, H$_2$CCC, is a very polar carbene, and the less stable isomer of $c$-C$_3$H$_2$ by $\sim$10~kcal~mol$^{-1}$ (5000\,K; \citealt{wu10}). Propadienylidene was first detected in the laboratory by \cite{vrt90} by means of millimetre-wave absorption spectroscopy in a DC discharge. It has been observed in prestellar and protostellar cores \citep{cer91, Turner00, Kawaguchi91,sak13}, circumstellar envelopes \citep{cer91}, translucent clouds \citep{Turner00}, diffuse clouds \citep{cer99,kul12,Liszt12}, and photo-dissociation regions (PDRs) \citep{Teyssier05, Cuadrado15}.

The cyclic to linear (hereafter $c/l$) ratio of C$_3$H$_2$ tends to increase with increasing $A_{\rm V}$. Significant variations of this ratio are found in dense cores, ranging from 25 in TMC-1(CP) to 67 in TMC-1C \citep{Spezzano16}. For completeness, we report in Table~\ref{tab0} the values of the $\rm C_3H_2$ $c/l$ ratio observed in several media, including dense cores. In order to be able to model the observed $c/l$ ratios, and to use this ratio as an astrochemical tool, the chemistry leading to the formation and destruction of $c$-C$_3$H$_2$ and H$_2$CCC (hereafter $l$-$\rm C_3H_2$) needs to be better understood, with emphasis on the branching ratios of the reactions involved.

The main formation process of cyclic and linear $\rm C_3H_2$ involves the radiative association of C$_3$H$^+$ and H$_2$ to form cyclic and linear C$_3$H$_3^+$, and a subsequent dissociative recombination with electrons that leads to $c$-C$_3$H$_2$ and $l$-$\rm C_3H_2$ respectively. C$_3$H$^+$ exists in both cyclic and linear forms, but the cyclic isomer lies 17\,kcal\,mol$^{-1}$ (8600\,K) above the linear one \citep{Ikuta97}, hence for our purposes we can assume to deal with $l$-C$_3$H$^+$ only. The radiative association of $l$-C$_3$H$^+$ and H$_2$ will form cyclic and linear C$_3$H$_3^+$ in a 1:1 ratio \citep{mal93}. \cite{tal09} suggest with their calculations that the dissociative recombination of $c$-C$_3$H$_3^+$  to form $c$-C$_3$H$_2$ is more efficient than its linear counterpart. Using an afterglow experiment at 300 K, \cite{ada05} observed that the cyclic C$_3$H$_3^+$  recombines faster than the linear isomer.

The chemistry of $c$ and $l$ $\rm C_3H_2$, and that of other associated species, has previously been modeled by, e.g., \citet{Turner00} and \citet{Fosse01}. Both of these works pointed out that also neutral-neutral reactions can play a part in determining the $\rm C_3H_2$ $c/l$ ratio. The rate coefficients of the reactions most relevant in the present context have changed a little since these studies.

A key point of the dissociative recombination of C$_3$H$_3^+$ is its branching ratio, which unfortunately has not been investigated yet neither experimentally nor theoretically. \cite{Chabot13} studied the dissociative recombination of cyclic and linear C$_3$H$_2^+$. Their results show that while $c$-C$_3$H is the main product of the dissociative recombination of $c$-C$_3$H$_2^+$ with electrons, $l$-C$_3$H is not the main product of the recombination of $l$-C$_3$H$_2^+$. In this work we study the effect of the branching ratios of the dissociative recombination of both cyclic and linear C$_3$H$_3^+$ on the cyclic to linear ratio of C$_3$H$_2$. We use the well-studied prestellar core L1544 to compare the results of our model with the observational results.

\section{Formation and destruction of $\rm C_3H_2$}\label{s:fd}

In physical conditions with large visual extinction ($A_{\rm V}$ > a few mag) where photoprocesses do not play a significant role, the formation and destruction of (cyclic and linear) $\rm C_3H_2$ follow a relatively simple scheme. The main formation pathway is the dissociative recombination $\rm C_3H_3^+ + e^-$. In this paper, we use the latest rate coefficient data from the KIDA database (\citealt{Wakelam15}; see also below), where three different branches for this reaction are included. These are tabulated in Table~\ref{tab1}. The third tabulated reaction for both forms of $\rm C_3H_2$, whose rate coefficient we label here as $k_{\rm DR,3}$, is the most important for the $\rm C_3H_2$ formation as it produces $\rm C_3H_2$ directly (see also Sect.\,\ref{ss:discconc}).

The main destruction partners of $\rm C_3H_2$ change with time, but at the time when $\rm C_3H_2$ is the most abundant (see Sects.~3 and 4), the main destruction reactions are ion-molecule reactions that convert $\rm C_3H_2$ back to $\rm C_3H_3^+$: $\rm C_3H_2 + HCO^+/H_3O^+/H_3^+ \longrightarrow C_3H_3^+ + CO/H_2O/H_2$ with rate coefficients $k_1$, $k_2$, and $k_3$, respectively. Thus, in steady-state, the abundance of $\rm C_3H_2$ would be given by
\begin{equation}
\Bigl[ {\rm C_3H_2} \Bigr] = \frac{ k_{\rm DR,3} \Bigl[ \rm C_3H_3^+ \Bigr] \Bigl[ \rm e^- \Bigr] } { k_1 \Bigl[ {\rm HCO^+} \Bigr] + k_2 \Bigl[ {\rm H_3O^+} \Bigr] + k_3 \Bigl[ {\rm H_3^+} \Bigr]} \, ,
\end{equation}
which leads to an expression of the $c/l$ ratio of $\rm C_3H_2$:
\begin{eqnarray}\label{c_to_l}
\frac{ \Bigl[ {c{\textrm -}\rm C_3H_2} \Bigr] }{ \Bigl[ {l{\textrm -}\rm C_3H_2} \Bigr] } &=& \frac{ k_{\rm DR,3}^c \Bigl[c{\textrm -}\rm C_3H_3^+ \Bigr] \Bigl[ \rm e^- \Bigr] }{ k_{\rm DR,3}^l \Bigl[ l{\textrm -}\rm C_3H_3^+ \Bigr] \Bigl[ \rm e^- \Bigr] } \nonumber \\
&\times& \frac{ k_1^l \Bigl[ {\rm HCO^+} \Bigr] + k_2^l \Bigl[ {\rm H_3O^+} \Bigr] + k_3^l \Bigl[ {\rm H_3^+} \Bigr]}{ k_1^c \Bigl[ {\rm HCO^+} \Bigr] + k_2^c \Bigl[ {\rm H_3O^+} \Bigr] + k_3^c \Bigl[ {\rm H_3^+} \Bigr]} \, .
\end{eqnarray}
Because of gas-grain chemical interaction, a steady-state solution is not attained in our models, but the above expressions turn out to give a rather accurate representation of the $\rm C_3H_2$ peak abundances. Inspection of the KIDA data reveals that the rate coefficients of the three $\rm C_3H_2$-destroying ion-molecule reactions discussed above are almost identical (slightly higher for $l$-$\rm C_3H_2$ than for $c$-$\rm C_3H_2$), so that the latter fraction in Eq.\,(\ref{c_to_l}) is $\sim$1. Furthermore, our chemical model (see below) indicates that the $c$-$\rm C_3H_3^+$ / $l$-$\rm C_3H_3^+$ ratio is usually also very close to unity, so that the $c$-$\rm C_3H_2$ / $l$-$\rm C_3H_2$ ratio is determined to a good approximation by the rate coefficient ratio $k_{\rm DR,3}^c / k_{\rm DR,3}^l$ at the time when $\rm HCO^+$, $\rm H_3O^+$, and $\rm H_3^+$ are abundant (i.e., before the onset of freeze-out of neutrals onto grain surfaces). This ratio is unity in the KIDA data which implies $c$-$\rm C_3H_2$ / $l$-$\rm C_3H_2 \sim 1$ according to the approximative formula presented above. However, the observed $c/l$ abundance ratio toward the dust peak in L1544 is $32 \pm 4$ \citep{Spezzano16}. We thus set out to modify the various $k_{\rm DR}$ branching ratios to explain the observations. The assumption that the two isomers are formed in the same fashion and in similar conditions is strengthened by our recent maps of $c$ and $l$ $\rm C_3H_2$ in the inner $2.5'\times2.5'$ of L1544 which demonstrate that $c$~and~$l$ $\rm C_3H_2$ trace the same region (Spezzano et al. in prep.). A spatial association of $c$~and~$l$ $\rm C_3H_2$ has been observed also in other sources (e.g. \citealt{Fosse01}).

\begin{table}
\caption{Branching ratios for $(c/l)-\rm  C_3H_3^+ + e^-$ from the KIDA database. The third column gives the denominations of the rate coefficients of the various branches discussed in the text. The total rate coefficient is in both cases $k_{\rm tot} = 7.0 \times 10^{-7} \, (T/300\,{\rm K})^{-0.5} \, \rm cm^3\,s^{-1}$.}
\centering
\begin{tabular}{c c c}
\hline \hline 
Reaction & Branching ratio & $k$\\ \hline
$c$-${\rm C_3H_3^+ + e^-} \longrightarrow \rm CH + C_2H_2 $ & 10\% & $k_{\rm DR,1}^c$ \\
$c$-${\rm C_3H_3^+ + e^-} \longrightarrow$ $c$-$\rm C_3H + H_2 $ & 45\% & $k_{\rm DR,2}^c$ \\
$c$-${\rm C_3H_3^+ + e^-} \longrightarrow$ $c$-$\rm C_3H_2 + H $ & 45\% & $k_{\rm DR,3}^c$ \\
\hline
$l$-${\rm C_3H_3^+ + e^-} \longrightarrow \rm CH + C_2H_2 $ & 10\% & $k_{\rm DR,1}^l$ \\
$l$-${\rm C_3H_3^+ + e^-} \longrightarrow$ $l$-$\rm C_3H + H_2 $ & 45\% & $k_{\rm DR,2}^l$ \\
$l$-${\rm C_3H_3^+ + e^-} \longrightarrow$ $l$-$\rm C_3H_2 + H $ & 45\% & $k_{\rm DR,3}^l$ \\
\hline
\end{tabular}
\label{tab1}
\end{table}

\begin{table}
\caption{Branching ratios calculated for $(c/l)$-$\rm C_3H_2^+ + e^-$ by \citet{Chabot13}. The total rate coefficient is in both cases $k_{\rm tot} = 4.2 \times 10^{-7} \, (T/300\,{\rm K})^{-0.5} \, \rm cm^3\,s^{-1}$.}
\centering
\begin{tabular}{c c}
\hline \hline 
Reaction & Branching ratio \\ \hline
$c$-$\rm C_3H_2^+ + e^- \longrightarrow C_2 + CH_2 $ & 3\% \\
$c$-$\rm C_3H_2^+ + e^- \longrightarrow H + H + C_3 $ & 13\% \\
$c$-$\rm C_3H_2^+ + e^- \longrightarrow H_2 + C_3 $ & 18\% \\
$c$-$\rm C_3H_2^+ + e^- \longrightarrow C + C_2H_2 $ & 27\% \\
$c$-$\rm C_3H_2^+ + e^- \longrightarrow {\rm H}$ + $c$-$\rm C_3H $ & 34\% \\
$c$-$\rm C_3H_2^+ + e^- \longrightarrow CH + CCH $ & 5\% \\
\hline
$l$-$\rm C_3H_2^+ + e^- \longrightarrow C_2 + CH_2 $ & 5\% \\
$l$-$\rm C_3H_2^+ + e^- \longrightarrow H + H + C_3 $ & 39\% \\
$l$-$\rm C_3H_2^+ + e^- \longrightarrow H_2 + C_3 $ & 4\% \\
$l$-$\rm C_3H_2^+ + e^- \longrightarrow C + C_2H_2 $ & 38\% \\
$l$-$\rm C_3H_2^+ + e^- \longrightarrow {\rm H}$ + $l$-$\rm C_3H $ & 7\% \\
$l$-$\rm C_3H_2^+ + e^- \longrightarrow CH + CCH $ & 7\% \\
\hline
\end{tabular}
\label{tab2}
\end{table}

\begin{figure*}
\centering
\resizebox{\hsize}{!}{\includegraphics[width=2.0\columnwidth]{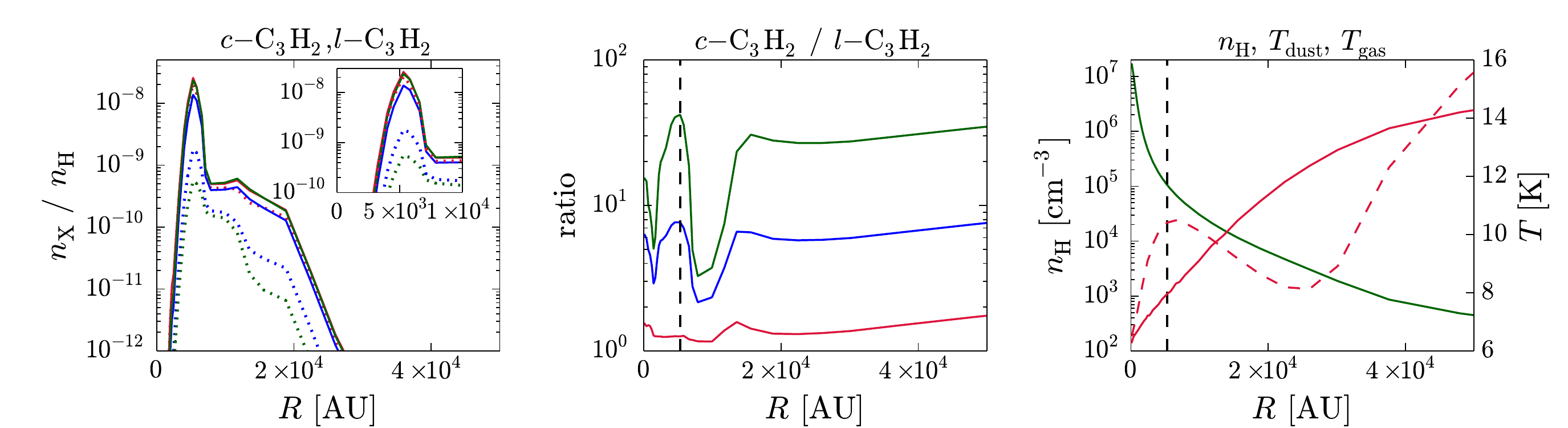}}
\caption{({\sl Left-hand panel}:) Abundances of $c$-$\rm C_3H_2$ (solid lines) and $l$-$\rm C_3H_2$ (dotted lines) as functions of distance from the core center in the L1544 model at $t = 10^5\, \rm yr$. ({\sl Middle panel}:) $\rm C_3H_2$ $c/l$ ratio as a function of distance from the core center in the L1544 model at $t = 10^5\, \rm yr$. ({\sl Right-hand panel}:) Density (green solid line), dust temperature (red solid line) and gas temperature (red dashed line) as functions of distance from the core center in the L1544 model \citet{Keto14}. The different colors in the left-hand and middle panels correspond to $k_{\rm DR,3}^c / k_{\rm DR,3}^l = 1$ (red), $\sim$7 (blue), and 25 (green). The inset in the left-hand panel shows a zoomed-in view of the innermost 10000~AU in the model core. The black vertical line marks the position of the $\rm C_3H_2$ abundance peak (the line is omitted in the left-hand panel for clarity of the figure).
}
\label{fig1}
\end{figure*}

As a guide to the modifications, we analyzed the branching ratios calculated for the $\rm C_3H_2^+ + e^-$ reaction by \citet{Chabot13}. These branching ratios, which are included in the KIDA database, are shown in Table~\ref{tab2}. Calculations for the $\rm C_3H_3^+ + e^-$ system are planned, but it is unclear when the results will be available (M. Chabot, priv. comm.). Therefore, we used the $\rm C_3H_2^+ + e^-$ system as a template for the modifications. In the dissociation of $\rm C_3H_2^+$, there are pathways that remove either one or both protons from $\rm C_3H_2^+$ and pathways that break C-C bonds, i.e., three groups of pathways. Inspection of the $\rm C_3H_3^+$ pathways (Table~\ref{tab1}) reveals a similar grouping. As a first approximation we combined the branching ratios belonging to each group in the $\rm C_3H_2^+$ data and translated them to the $\rm C_3H_3^+$ system. This leads to the branching ratios given in Table~\ref{tab3} (mod\_ch). In this modified model, the $k_{\rm DR,3}^c / k_{\rm DR,3}^l$ ratio is $\sim$5, which is still not high enough in light of the observations. We then proceeded to vary the branching ratios to attain a fit to the observed $\rm C_3H_2$ $c/l$ ratio in L1544. The results of our analysis are presented in Sect.\,\ref{ss:results}.

\begin{table}
\caption{Modified branching ratios for $(c/l)$-$\rm C_3H_3^+ + e^-$ based on the $(c/l)$-$\rm C_3H_2^+ + e^-$ data from \citet{Chabot13} (labeled mod\_ch), and an ad hoc approach (mod\_hr; ``hr'' stands for ``high ratio'').}
\centering
\begin{tabular}{c c c}
\hline \hline 
Reaction & mod\_ch & mod\_hr \\ \hline
$c$-${\rm C_3H_3^+ + e^-} \longrightarrow \rm CH + C_2H_2 $ & 35\% & 25\%\\
$c$-${\rm C_3H_3^+ + e^-} \longrightarrow$ $c$-$\rm C_3H + H_2 $ & 31\% & 25\%\\
$c$-${\rm C_3H_3^+ + e^-} \longrightarrow$ $c$-$\rm C_3H_2 + H $ & 34\% & 50\%\\
\hline
$l$-${\rm C_3H_3^+ + e^-} \longrightarrow \rm CH + C_2H_2 $ & 50\% & 49\%\\
$l$-${\rm C_3H_3^+ + e^-} \longrightarrow$ $l$-$\rm C_3H + H_2 $ & 43\% & 49\%\\
$l$-${\rm C_3H_3^+ + e^-} \longrightarrow$ $l$-$\rm C_3H_2 + H $ & 7\% & 2\%\\
\hline
\end{tabular}
\label{tab3}
\end{table}

To simulate the $\rm C_3H_2$ abundances in L1544, we used the physical model for L1544 presented by \citet{Keto10} and updated by \citet{Keto14} which gives us the density, (gas and dust) temperature, and $A_{\rm V}$ profiles as functions of distance away from the dust peak. We separated the model core to concentric shells and calculated the chemical evolution separately in each shell. In this way we can produce simulated chemical abundance gradients and study how the column densities of the various molecules change as functions of time. A similar procedure has been used to model successfully the emission/absorption of the ortho and para $\rm H_2D^+$ ground-state rotational lines toward IRAS~16293A \citep{Brunken14}. The (gas-grain) chemical code used here is described in detail in \citet{Sipila15a}. The gas-phase chemical reaction set used in the present work is based on the latest KIDA reaction file \citep{Wakelam15}, which was deuterated and spin-state separated according to the prescriptions of \citet{Sipila15a} and \citet{Sipila15b}. However, in this paper we do not explicitly consider spin states or deuteration. The physical parameters of the model and the initial chemical abundances are adopted from Tables~1 and 3 in \citet{Sipila15a} except for the diffusion to binding energy ratio ($E_{\rm d} / E_{\rm b}$) for which we adopt here a value of 0.60, in line with the recent results of \citet{Minissale16}.

\section{Results}\label{ss:results}

We calculated the abundances and column densities of $c$-$\rm C_3H_2$ and $l$-$\rm C_3H_2$ in the L1544 model with a grid of values for the $k_{\rm DR,3}^c / k_{\rm DR,3}^l$ ratio, ranging from 1/1 (the KIDA value) up to 50/2 (mod\_hr, Table~\ref{tab3}). We plot in Fig.\,\ref{fig1} the abundances of $c$-$\rm C_3H_2$ and $l$-$\rm C_3H_2$ and the $\rm C_3H_2$ $c/l$ ratio at $t = 10^5$\,yr assuming three different values for $k_{\rm DR,3}^c / k_{\rm DR,3}^l$. We also show for completeness the density and temperature structures of the L1544 core model \citep{Keto14}. Additional calculations (not shown) indicate a simple increasing trend of the $\rm C_3H_2$ $c/l$ ratio with $k_{\rm DR,3}^c / k_{\rm DR,3}^l$. Evidently, neither the KIDA model nor our initial modified model (mod\_ch) produce a high enough $c/l$ ratio, and a high $k_{\rm DR,3}^c / k_{\rm DR,3}^l$ is needed to produce a $c/l$ ratio comparable to the observed value.

\begin{figure*}
\centering
\resizebox{\hsize}{!}{\includegraphics[width=2.0\columnwidth]{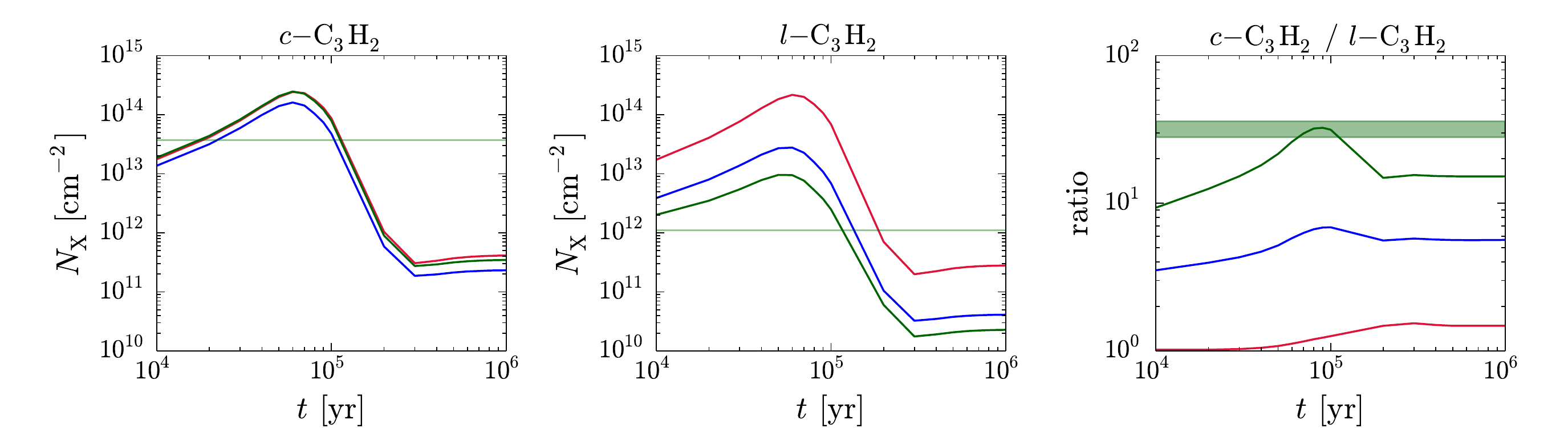}}
\caption{Column densities (convolved to a $30''$ beam) of $c$-$\rm C_3H_2$ (left panel) and $l$-$\rm C_3H_2$ (middle panel), and the $\rm C_3H_2$ $c/l$ column density ratio (right panel) as functions of time toward the center of the L1544 model. The different colors correspond to $k_{\rm DR,3}^c / k_{\rm DR,3}^l = 1$ (red), $\sim$7 (blue), and 25 (green). The horizontal lines/bars in each panel correspond to the observed values toward the L1544 dust peak \citep{Spezzano16}. 
}
\label{fig2}
\end{figure*}

To supplement the results shown in Fig.\,\ref{fig1}, we show in Fig.\,\ref{fig2} the column densities of cyclic and linear $\rm C_3H_2$ and their ratio as functions of time toward the center of the L1544 model. We also show the observed values \citep{Spezzano16}. The simulated column densities were convolved to a $30''$ beam so that they correspond to the observations. The peak value of the $\rm C_3H_2$ $c/l$ ratio is $\sim$32 which corresponds to $8 \times 10^4$\,yr of chemical evolution. The best fit to the observations is attained at $t \sim 10^5$\,yr, when the column densities of both species and their ratio are all within the observed limits. The timescale required to attain the best fit depends on the initial chemical abundances and so our results do not necessarily reflect the ``absolute'' age of the core. However, it is clear that we can only attain a high enough $\rm C_3H_2$ $c/l$ ratio with the model if the production of cyclic $\rm C_3H_2$ is strongly boosted over that of linear $\rm C_3H_2$ in the dissociation of $\rm C_3H_3^+$.

\section{Discussion and conclusions}\label{ss:discconc}

Looking at the $\rm C_3H_3^+ + e^-$ branches (Tables~\ref{tab1}~and~\ref{tab3}), we see that the branch listed second ($\rm C_3H_3^+ + e^- \longrightarrow C_3H + H_2$) forms $\rm C_3H$, which can be converted back to $\rm C_3H_2$ through the sequence $\rm C_3H + H^+ \longrightarrow C_3H^+ + H$; $\rm C_3H^+ + H_2 \longrightarrow C_3H_3^+$; $\rm C_3H_3^+ + e^- \longrightarrow C_3H_2 + H$. Therefore it seems reasonable to assume that increasing $k_{\rm DR,2}^c / k_{\rm DR,2}^l$ would further increase the $\rm C_3H_2$ $c/l$ ratio. We tested this by modifying the $k_{\rm DR,2}^c / k_{\rm DR,2}^l$ and $k_{\rm DR,1}^c / k_{\rm DR,1}^l$ ratios, keeping $k_{\rm DR,3}^c / k_{\rm DR,3}^l$ fixed at 50/2. We found that increasing $k_{\rm DR,2}^c$ and decreasing $k_{\rm DR,2}^l$ by 10\%\footnote{By this we mean an increase/decrease of the branching ratio by 10 units of percentage, as opposed to 10\% of the original value.} further increases the peak $\rm C_3H_2$ $c/l$ ratio to $\sim$40, while decreasing $k_{\rm DR,2}^c$ and increasing $k_{\rm DR,2}^l$ by 10\% leads to a peak $\rm C_3H_2$ $c/l$ ratio of $\sim$26. Therefore we can attain a good fit to the observations with several combinations of the $k_{\rm DR,2}$ and $k_{\rm DR,3}$ values, and it is difficult to derive strict upper limits to ``acceptable'' values of the branching ratios based on the present analysis. However, our calculations indicate that the effect of $\rm C_3H_3^+ + e^- \longrightarrow C_3H_2 + H$ is greater than that of $\rm C_3H_3^+ + e^- \longrightarrow C_3H + H_2$ in determining the $\rm C_3H_2$ $c/l$ ratio. This is because the former pathway is a direct source of $\rm C_3H_2$ while the latter pathway depends on multiple reactions (see above) that can also form species other than $\rm C_3H_2$. We require $k_{\rm DR,3}^c / k_{\rm DR,3}^l$ clearly higher than unity. We note that the $\rm C_3H$ $c$/$l$ ratio is in our models nearly independent of the various $k_{{\rm DR},n}$ values, and is $\sim$3 at the time of the $\rm C_3H_2$ peak, increasing to $\sim$10 at later times (as the $\rm C_3H_2$ $c$/$l$ ratio goes down). These values are in agreement with observations (\citealt{Turner00}, \citealt{Fosse01}, \citealt{Cuadrado15}). However, because of its weak dependence on $k_{{\rm DR},n}$, the $\rm C_3H$~$c$/$l$ ratio cannot be used to constrain the branching ratios of the $\rm C_3H_3^+ + e^-$ process.

Figure~\ref{fig2} demonstrates that the $\rm C_3H_2$ $c/l$ ratio is not constant in time, i.e., that the approximation of Eq.\,(\ref{c_to_l}) does not hold universally. For the highest $k_{\rm DR,3}^c / k_{\rm DR,3}^l$ ratio tested here (50/2, Table~\ref{tab3}), the $\rm C_3H_2$ $c/l$ ratio presents a maximum for a relatively brief period of time and finally settles to $\sim$15. Our chemical model shows that at late times the slow neutral-neutral reaction $\rm H + CH_2CCH \longrightarrow H_2 + C_3H_2$ is a significant source of $\rm C_3H_2$. In fact, because the $k_{\rm DR,3}^c / k_{\rm DR,3}^l$ ratio is so high, this neutral-neutral reaction is the most important formation pathway for $l$-$\rm C_3H_2$ (and the second most important pathway for $c$-$\rm C_3H_2$), so that the $\rm C_3H_2$ $c/l$ ratio is no longer controlled solely by $\rm C_3H_3^+ + e^-$. This finding is in line with previous discussions by \citet{Turner00} and \citet{Fosse01}.

From Fig.\,\ref{fig1} we see that the $\rm C_3H_2$ $c/l$ abundance ratio presents a minimum about 10000~AU away from the core center, and increases again at larger radii. This behavior is a result of changing physical conditions, particularly the visual extinction. A detailed analysis of our results reveals that around 10000~AU at $t = 10^5\,\rm yr$, the formation of linear $\rm C_3H_2$ is governed by reactions other than $\rm C_3H_3^+ + e^-$ because of its low rate coefficient, while at still higher radii photochemistry comes into play and the $\rm C_3H_3^+ + e^-$ reaction is again the dominant formation pathway for both cyclic and linear $\rm C_3H_2$. However, the abundances are very low in the outer core; by far the largest fraction of the cyclic and linear $\rm C_3H_2$ column densities comes from the interval $\sim$3000--7000\,AU.

The total rate coefficient for $\rm C_3H_3^+ + e^-$ assumed in the earlier modeling works of \citet{Turner00} and \citet{Fosse01} is equal to that adopted here (Table~\ref{tab1}) within a factor of two. Although only a limited amount of rate coefficient data is quoted in \citet{Turner00} and \citet{Fosse01}, it seems that the rate coefficients for the main reactions in the present context have not significantly changed in the last two decades. Still, it is clear from all the models that the observed $c$/$l$ ratios can only be reproduced if the relative rate coefficients for the formation and destruction of the various $c$ and $l$ species have different values.

As stated in the Introduction and expanded upon in Appendix~\ref{appxa}, the $\rm C_3H_2$ $c/l$ ratio has been observed toward various environments (e.g., protostellar cores, diffuse clouds) with values ranging from 3 to $\sim$70. The observations show a trend of increasing $\rm C_3H_2$ $c/l$ ratio at increasing $A_{\rm V}$. The analysis presented above demonstrates that various reactions can be responsible for the formation of $\rm C_3H_2$ (and hydrocarbons in general; see \citealt{Alata14,Alata15}; \citealt{Duley15}). We cannot, based on the present models, draw quantitative conclusions on the expected $\rm C_3H_2$ $c/l$ ratio in objects other than L1544 without detailed (physical) modeling, particularly toward environments with low gas density and low $A_{\rm V}$. Such an undertaking is beyond the scope of this Letter. As an example of the interpretational difficulty, we note that the $\rm C_3H_2$ $c/l$ ratio in the outer parts of the L1544 model, which is roughly consistent with a PDR in terms of physical conditions, follows the $k_{\rm DR,3}^c / k_{\rm DR,3}^l$ ratio (Fig.\,\ref{fig1}) and we checked that the $\rm C_3H_2$ $c/l$ column density ratio is indeed high ($\sim$17--35 depending on the time) even 100" away from the core center. This is in agreement with the observations (Spezzano et al. in prep.).

We conclude that in order to explain the $\rm C_3H_2$ $c/l$ ratio observed toward L1544 ($32 \pm 4$; \citealt{Spezzano16}), the branching ratios of the various outcomes of the $\rm C_3H_3^+ + e^-$ reaction should be adjusted to favor cyclic $\rm C_3H_2$ over linear $\rm C_3H_2$. In physical conditions with moderate visual extinction, the most important reaction determining the $\rm C_3H_2$ $c/l$ ratio is $\rm C_3H_3^+ + e^- \longrightarrow C_3H_2 + H$, followed by $\rm C_3H_3^+ + e^- \longrightarrow C_3H + H_2$. It is currently assumed in the KIDA database that $c$ and $l$ $\rm C_3H_2$ are created in equal proportions in both the former and the latter reaction. However, our modeling results suggest that the ratio of the rate coefficients in the former reaction for cyclic and linear $\rm C_3H_2$ should be of the order of $\sim$20 depending on what is assumed for the latter reaction. Unfortunately, we cannot set strict limits on the branching ratios based on our current analysis. More theoretical and/or laboratory work is clearly needed to understand the chemistry of $\rm C_3H_2$.

\begin{acknowledgements}

We thank the anonymous referee for helpful comments and suggestions that improved the manuscript. We acknowledge the financial support of the European Research Council (ERC; project PALs 320620).

\end{acknowledgements}

\bibliographystyle{aa}
\bibliography{c3h2.bib}

\onecolumn 

\begin{appendix}

\section{Additional data}\label{appxa}

As noted in the introduction, the C$_3$H$_2$ $c/l$ ratio tends to increase with increasing $A_{\rm V}$. Table~\ref{tab0} reports the values of the $c/l$ ratio observed in several media. The values range from 3 in diffuse clouds to over 50 in dense cores. However, the density of the medium is not the only quantity that has an effect on this ratio. \cite{Teyssier05} showed that inside the Horsehead nebula PDR the C$_3$H$_2$ $c/l$ ratio  does not exceed 15 despite the extinction being of the order of 10-20, which is comparable with what is found in TMC-1(CP), where the $c/l$ ratio has been observed to be 25 \citep{Turner00}. In the Orion Bar a $c/l$ ratio of 34 has been recently observed \citep{Cuadrado15}, which is larger than the value of $\sim$4 observed towards the Horsehead PDR \citep{Teyssier05}. Furthermore, the $c/l$ ratio is also different among dense cores, ranging from 25 in TMC-1(CP) to 67 in TMC-1C \citep{Spezzano16}.

\begin{table*}[h]
\caption{Observed $\rm C_3H_2$ $c/l$ ratios toward various objects, and the associated physical conditions (visual extinction $A_{\rm V}$, medium density $n$, (rotational) temperature $T$) when available in the literature.}
\centering
\begin{tabular}{cccccc}
\hline\hline
Source & $A_{\rm V} \, [\rm mag]$ & $n \, [\rm cm^{-3}]$ & $T_{\rm rot}^{(a)} \, [\rm K]$ & $c/l$ ratio & Reference\\
\hline\hline
\textbf{Prestellar cores}&&&&&\\
TMC-1(CP$^{(b)}$)&&$3\times10^4$& 7 &70&\cite{cer91}\\
TMC-1(CP)&&$3\times10^4$& 7 &25&\cite{Turner00}\\
TMC-1(CP)&&$3\times10^4$& 7 &20-40&\cite{Kawaguchi91}\\
TMC-1(CP)&&$3\times10^4$& 7 &28&\cite{Fosse01}\\
TMC-1(edge$^{(c)}$)&&$3\times10^4$& 7 &10&\cite{Fosse01}\\
L183&&&& 50&\cite{Turner00}\\
TMC-1C&&& 7 ($c$), 4 ($l$) & 67&\citet{Spezzano16}\\
L1544&&& 6 ($c$), 4 ($l$) & 32 &\citet{Spezzano16}\\
\textbf{Protostellar cores}&&&&&\\
L1527& &$7\times10^5$ & 12 &12&\cite{sak13}\\
\textbf{Translucent clouds}&&&&&\\
CB17, CB24, CB228 &&&&17(average)&\cite{Turner00}\\
\textbf{Circumstellar envelopes}&&&&&\\
IRC+10216&&& 25 &30&\cite{cer91}\\
\textbf{Diffuse clouds}&&&&&\\
W49N&&& $T_{\rm CMB}$$^{(d)}$ &4&\cite{kul12}\\
W51&& & $T_{\rm CMB}$ &13&\cite{kul12}\\
B0355+508, B0415+379,&&& $T_{\rm CMB}$ & 15-40 & \citet{Liszt12}\\
B2200+420, B2251+158 &&&& & \\
\textbf{PDRs}&&&&&\\
Horsehead Nebula (edge$^{(e)}$) & & $< 10^{4\,(f)}$ &&3-5&\cite{Teyssier05}\\
Horsehead Nebula (dense cloud$^{(e)}$)& 10-20 & $2 \times 10^{5\,(f)}$ &&15&\cite{Teyssier05}\\
Orion Bar &&& 17 ($c$), 26 ($l$) & 34 &\citet{Cuadrado15}\\
\hline
\end{tabular}
\tablefoot{$^{(a)}$ The rotational temperature of $c$ and/or $l$ $\rm C_3H_2$, which may not be equal to the kinetic temperature. $^{(b)}$ CP stands for Cyanopolyyne Peak. $^{(c)}$ The edge position is (-40,0) offset from the CP position $\alpha_{2000} = 4^{\rm h} \, 41^{\rm m} \, 42.5^{\rm s}$, $\delta_{2000} = 25^{\circ} \, 41'\, 27''$ \citep{Kawaguchi91}. $^{(d)}$ $T_{\rm CMB}$ stands for the temperature of the Cosmic Microwave Background (2.7\,K). $^{(e)}$ The edge and cloud positions are (by estimation) (-20,-15) and (-72,-62) offset from the (0,0) position \citep{Teyssier05}. $^{(f)}$ Density corresponds to the model presented by \citet{Pety12}.
}
\label{tab0}
\end{table*}

\end{appendix}

\end{document}